\documentclass[11pt,twoside]{article}

\usepackage{asp2006}
\usepackage{epsf}

\usepackage{lscape}

\begin{document}
\title{E+S galaxy pairs: 
are they the precursors of fossil groups?}   %%% Fill in title
\author{R. Gr\"utzbauch$^{1,2}$, F. Annibali$^{3}$, R. Rampazzo$^{3}$, W. W. Zeilinger$^{2}$}   %%% Fill in author names
\affil{$^{1}$School of Physics and Astronomy, Univ. of Nottingham, UK \\
$^{2}$Institute for Astronomy, Univ. of Vienna, Austria \\
$^{3}$INAF – Osservatorio Astronomico di Padova, Italy \\}     %%% Fill in author affiliations

~\\

Galaxy pairs may represent a way station in the evolutionary path from poor groups to giant isolated ellipticals. To test this evolutionary scenario, we investigated the environment of 4 galaxy pairs composed of a giant elliptical galaxy (E) and its spiral companion (S). The pairs are very similar from the optical and dynamical point of view, but have very different X-ray properties. Two of them show extended diffuse X-ray emission from a hot Intra Group Medium (IGM), whereas the other two seem to be deficient in hot gas (Gr\"utzbauch et al. 2007, AJ 133, 220).

\noindent
Can this difference be caused by the environment of the galaxy pair or is it indicative of a different evolutionary phase of the pair or the elliptical? Are these pairs a way station in the coalescence process of galaxy groups towards isolated ellipticals with extended group-like IGM, the famous fossil groups? \\

\noindent
We investigated the group dynamics, the group member's morphologies and luminosity function (LF), separated into X-ray bright and X-ray faint groups.\\

\noindent
{\bf group dynamics:}
We find that our pairs are the dominating members of most probably bound structures as suggested by the group members' short crossing times and their morphology density relation.\\

\noindent
{\bf LFs:}
Our {\bf X-ray bright} LF is not consistent with any of the literature LFs (Lin et al. 1996, ApJ 464, 60; Zabludoff \& Mulchaey 2000, ApJ 539, 136; Mendes de Oliveira et al. 2006, AJ 131, 158) and could indicate that our X-ray bright systems represent a more evolved stage in group evolution.
Our {\bf X-ray faint} LF resembles the LF of X-ray bright groups from the literature and may represent a phase in the dynamical evolution of these groups, where the recent or ongoing interaction, in which the pair E is involved, has destroyed or at least decreased the luminosity of the IGM. The X-ray faint groups' LF is also consistent with their evolution into a fossil group. \\

\noindent
{\bf ages and metallicities:}
 The ongoing study of absorption line indices of the group members indicates a more gradual chemical enrichment history for our dwarfs than for giant early-type galaxies.\\

\noindent
The diverse properties of our E-dominated poor groups indicate that the process of fossil group formation may not be as straightforward as commonly thought.

%%% MAIN BODY OF TEXT GOES HERE. CONSULT "INSTRUCTIONS FOR AUTHORS USING
%%% LATEX2E MARKUP", SECTIONS 2.3-2.6 FOR HELP WITH EQUATIONS, FIGURES,
%%% AND TABLES.

%\section{}   %%% Top level section head (remove "%" symbol)
%\subsection{}   %%% Second level section head (remove "%" symbol)
%\subsubsection{}   %%% Lowest level section head (remove "%" symbol)
%\section*{}    %%% Unnumbered top level section head (remove "%" symbol)
%\subsection*{}   %%% Unnumbered second level section head (remove "%" symbol)

%%% THE BIBLIOGRAPHY
%%%
%%% CONSULT SECTION 3 OF "INSTRUCTIONS FOR AUTHORS" FOR HOW TO USE NATBIB.
%%% AUTHORS ARE ENCOURAGED TO USE EITHER THE "THEBIBLIOGRAPY" ENVIRONMENT
%%% BY UNCOMMENTING (DELETING THE "%" SYMBOL) THE COMMANDS BELOW, OR BY
%%% USING THE BIBTEX ENVIRONMENT. TO FIND OUT WHICH IS APPLICABLE TO YOUR
%%% CONTRIBUTION, CONSULT THE VOLUME EDITORS FOR YOUR PROCEEDINGS.
%%%

%\begin{thebibliography}{}
%\bibitem[]{}
%\bibitem[]{}
%\bibitem[]{}
%\bibitem[]{}
%\bibitem[]{}
%\bibitem[]{}
%\bibitem[]{}
%\bibitem[]{}
%\bibitem[]{}
%\bibitem[]{}
%\bibitem[]{}
%\bibitem[]{}
%\end{thebibliography}

\end{document}